\documentclass{article}
\usepackage[top=3cm, bottom=3cm, left = 3cm, right = 3cm]{geometry} 
\usepackage{amsmath,amssymb}  
\usepackage{bm}  
\usepackage{booktabs}
\usepackage{caption}
\usepackage{fancyhdr}
\usepackage{float}
\usepackage{graphicx} 
\usepackage[utf8]{inputenc}
\usepackage{memhfixc} 
\usepackage{multirow}
\usepackage{placeins}
\usepackage{textcomp}
\usepackage[pdftex,bookmarks,colorlinks,breaklinks]{hyperref}  
\usepackage[style=apa]{biblatex}
\newcommand{\keywords}[1]{\par\noindent\textbf{Keywords: }#1}

\begin{document}
 \title{Preprints Without Curation Are Increasingly Cited by Journals}
\author{
 Chiaki Miura$^{1a*}$, Ichiro Sakata$^{a}$\\
    \small $^{1}$ORCID: 0009-0009-6492-0985 \\
    \small $^{a}$Department of Engineering, Technology Management and Innovations,\\
    \small The University of Tokyo, Bunkyo, Tokyo 113-0033, Japan  \\
    \small $^{*}$Corresponding author: \tt{miura-tchiaki873@g.ecc.u-tokyo.ac.jp} \\
}
\date{This manuscript was compiled on \today}

\maketitle 
\begin{abstract}
The recent strain on peer review systems necessitates the publish–review–curate model, where the public dissemination of manuscripts is decoupled from formal peer review and later editorial processes.
Citation norms traditionally pointed to the \textit{curated} version; those that subsequently appear in peer-reviewed journals.Citations to preprints are therefore generally expected to occur only to those eventually curated.
Using metadata from six major preprint servers linked to the world's largest bibliographic reference database, and focusing on citations from long-established journals with rigorous editorial workflows, we compare observed citation rates with a null model reflecting the expected rate if authors were indifferent between preprints and journal articles.
The growth is evident in both citation intensity and the breadth of adoption: at its peak in 2021, nearly one in five journal articles cited at least one preprint.
The preference to cite preprints has grown exponentially since the mid-2010s, with key inflection points coinciding with the launch of bioRxiv (2013) and the introduction of a preprint content type in Crossref's DOI system (2016).
An increasing share of these citations refer to non-curated preprints and this shift cannot be explained by confounding factors: the growing stock of preprints, faster publishing cycles, or the outsized influence of a small set of highly cited preprints.
These findings reveal a structural shift in citing behavior: preprints are increasingly treated as legitimate, autonomous objects of citation, independent of their later publication status, motivating renewed attention to how citation conventions and journal policies should accommodate the expanding role of preprints in scholarly communication.
Contrary to expectations, the COVID-19 pandemic did not amplify this trend but instead coincided with a deceleration in preprint citation growth.
\end{abstract}

\section{Introduction}
On December 7, 2000, Nature published a paper titled ``Quantum theory's last challenge''\parencite{Amelino-Camelia_QuantumTheorysLast_2000}.
The paper is one among many quantum physics papers, but it was the first article in the journal to explicitly cite an arXiv without labeling it as a preprint.
Although many researchers, particularly astronomers, habitually cite articles on arXiv (and its predecessor, the Los Alamos National Laboratory e-print server), such citations had previously been accompanied by annotations such as "submitted", "accepted", and "in print", often including phrases like "available at \textlangle preprint URL\textrangle".
In this instance, however, the arXiv article was never submitted to any journal, and stayed as the single legitimate source of knowledge.
It was followed by other examples: a series of research articles in 2002 to 2003 by Gregori Perelman, which was published only on arXiv and the three preprints combining to get cited over 6,000 times as of 2025. 

Then the question arises: How common is it for preprints to be cited as the sole source of knowledge in other disciplines? Does it frequently occur only in preprint friendly fields? Does this phenomenon reflect a change in the role of preprints, or is it the accelerated research cycles that urged researchers to cite a preliminary work?
The answers to these questions have significant implications for scholarly communication, as preprints bypass the traditional peer-review process that has long underpinned the credibility of scientific knowledge.

Preprints have first established its role in fast dissemination.
Preprints are intended to minimize the publishing delay due to article processing\parencite{Goldschmidt-Clermont_CommunicationPatternsHighEnergy_2002}.
The founder of the largest preprint server arXiv, Paul Ginsparg, has noted this evolution.
What was once used mainly in physics and mathematics extended into computer science and later expanded into biology\parencite{BergEtAl_PreprintsLifeSciences_2016} and psychology.
Researchers consider posting preprints for diverse objectives; maximizing impact\parencite{KurtzEtAl_EffectUseAccess_2005}, open access, claiming priority, and early feedback.
Yet, the primary motivations has been to ``publicize my research as soon as I think it is ready,'' consistently across major STEM fields\parencite{FraserEtAl_MotivationsConcernsSelection_2022, NiWaltman_PreprintNotPreprint_2024, Kelly_SIGIRCommunitySurvey_2018}

However, over the past three decades the preprint landscape has transformed dramatically: 
three disciplines accounted for the vast majority (93\%) of arXiv submissions in 2010–2011: mathematics (with 21\% of all WoS papers on arXiv), physics (20\% of all WoS papers on arXiv) and earth and space (12\% of all WoS papers on arXiv) \parencite{LariviereEtAl_ArXivEprintsJournal_2014}.
While the integrated use of preprints become prominent in computer science, the late 2010s witnessed the launch of new preprint servers centered on the Open Science Framework (OSF), including SocArXiv (2016) for social sciences, PsyArXiv (2016) for psychology, and EngrXiv (2017) for engineering.
The emergence and expansion of preprint review platforms, funder policy changes and community advocacy for preprint adoption continues to support this change.
As the post-publication peer review pipeline matures in processing and verifying more articles on preprint servers\parencite{WeissgerberEtAl_AutomatedScreeningCOVID19_2021}, emergent publication platforms such as eLife and F1000 consider preprints as an independent, primary medium of academic discourse.
Although preprints contribute to accelerating science, researchers still debate whether they should treat them as equivalent to traditional publication channels.

Despite these transformations, the role of preprints in the production and circulation of knowledge remains poorly understood.
Most empirical studies still treat journal articles as the primary unit of analysis and examine preprints mainly through their relationship to the journal “version of record”. 
For example, prior work has compared preprints with their subsequently published versions and found only minor differences in authorship and references \parencite{AkbaritabarEtAl_StudyReferencingChanges_2022}, as well as broadly similar expert evaluations \parencite{BeroEtAl_ComparisonPreprintsFinal_2021}.
Few exceptions have examined the number of citations directly to preprints, finding that preprints are cited in the first few months after posting, and decline after journal publication \parencite{FraserEtAl_RelationshipBioRxivPreprints_2020}

While these comparisons are valuable, they implicitly frame preprints as preliminary versions of journal articles, overlooking their independent role in scholarly communication. As preprints increasingly circulate, are evaluated, and are cited prior to formal publication, their role cannot be understood solely through comparison with their later journal counterparts. A more direct approach is to examine how preprints are actually used within the scientific literature, and citation patterns offer the clearest window into this process.

In recent years, scholarly communication has increasingly shifted toward a publish–review–curate model\parencite{EisenEtAl_ImplementingPublishThen_2020} in which the initial dissemination of a manuscript (the “publish” stage) is decoupled from both formal peer review and its later evaluation within the research community. 
Within this framework, the transition from a publicly posted preprint to a journal article can be understood less as a simple act of publication and more as a process of curation. 
Curation denotes the selective, evaluative, and editorial processes through which certain works are highlighted, assessed, and stabilized within scholarly communication.
Accordingly, we refer to preprints that later appear in peer-reviewed journals as \textit{curated} preprints.

To examine how preprints become integrated into ongoing scientific discourse, we focus on citations to preprints from journal articles. Our aim is not to treat journal publication as the ultimate marker of scholarly legitimacy. 
Rather, journal articles provide a particularly informative observational window into how researchers actively connect pieces of knowledge when constructing new work. 
Because references represent deliberate decisions by authors about which prior results to incorporate, they reveal how preprints are taken up in the production of subsequent research in the tension between formal validation and community endorsement.

Using the world’s largest bibliographic reference dataset combined with server-level metadata from major preprint repositories, we quantify the extent to which preprints are cited by journal articles while distinguishing between curated and non-curated preprints. 
We focus empirically on biology, where the expansion of preprint adoption has been particularly pronounced, providing a clear setting to examine how the role of preprints is evolving within contemporary scholarly communication.

\section{method}
\subsection{Preprint criteria}
Preprint is a complete scientific manuscript that is publicly shared by the authors on a dedicated repository which have never recieved any formal peer review.
More precise definition of preprint is still debated, but after \cite{BergEtAl_PreprintsLifeSciences_2016}, we operationally delimit preprint server with the following criteria:
1. The server is dedicated to research output that are complete, not peer reviewed articles;
2. A sizable amount of articles have been posted on the server and is not the result of curating from other sources;
3. All or most of the articles are free to post and freely accessible, both for majority of researchers;
4. Complete metadata, including co-authors, title, reference and consistent uid (especially DOI), is available.
The candidate preprint servers are those listed in the Directory of Open Access Preprint Repositories (DOAR)\parencite{_DirectoryOpenAccess_}.
DOAR lists 92 preprints servers as of November 2025.

We set the size threshold of 1,000 articles in total in the second criteria since most functioning repositories host more than 1,000 articles \parencite{Bjork_OpenAccessSubject_2014}.
We added the third criteria because the access privilege of author and audience may affect the citation behavior, and thus structural difference between open and close preprint servers should be avoided.

In total six preprint servers met the criteria: arXiv, bioRxiv, medRxiv, OSF Preprints, SocArXiv and PsyArXiv.
Mathematical Physics Preprint Archive (mp\_arc), LingBuzz, Cryptology ePrint Archive, PhilArchive, and ViXra are also satisfying the criteria except they only retain their own unique identifier system, and do not provide DOIs. 
The Web of Science preprint citation index, Europe PMC, and preprint citation indexes handle only preprint servers that provide DOIs.
None of the reliable bibliographic database indexes articles on the servers.
The absence of DOIs on some servers imposes an inherent limitation:
preprints without DOIs are generally not re-indexed by major bibliographic databases, which makes it technically impossible to track their citations in a systematic manner.
Assigning DOIs can improve their discoverability and, in turn, the likelihood of being cited.
We cover this issue in the discussion section.

\subsection{Data Acquisition}\label{data_acquisition}
Preprint metadata are obtained from their respective servers or partnered repositories. 
ArXiv data is obtained from the Zenodo arXiv datadump \parencite{Geiger_ArXiVArchive_2019}. 
The dataset is a complete metadata from 1993-2018 via Open Archives Initiative Protocol for Metadata Harvesting (OAI-PMH), and we complemented the recent (2019-2024) data via official arxiv API.
BioRxiv and medRxiv data is obtained via bioRxiv and medRxiv API provided by the Cold Spring Harbor Laboratory.
OSF Preprints, SocArXiv, PsyArXiv data is obtained via OSF API.

For the reference relationships, we used the world largest biblographic database, OpenAlex. 
OpenAlex is an open bibliographic database that covers over 240 million scholarly works across various disciplines.
We used the bulk snapshot available at 1. May, which may include the most of the updates up to the end of 2024, considering several month delay from the publishers and data-harvesters.
We matched the preprint metadata to OpenAlex records by DOI.
In total 3,015,656 preprints were matched to OpenAlex records out of 3,084,686 preprints obtained, 1,905,278 of these were curated preprints.
Detailed matching rates by preprint servers are shown in Table~\ref{tab:preprint_matching}.

To assess whether the observed citation trends hold after removing the substantial shifts in publication patterns during the pandemic, we identified COVID-19-related journal articles as those containing ``covid,'' ``covid-19,'' or ``sars-cov-2'' in their title or abstract.

\subsection{Curation of Preprints} \label{sec:methods_preprint_curation}
\begin{figure}[H]
  \centering
  \includegraphics[width=0.5\linewidth]{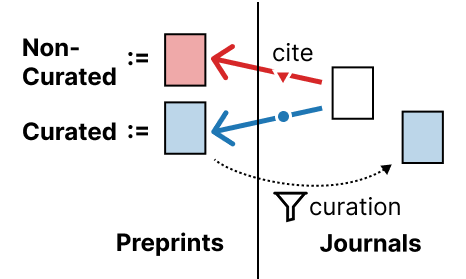}
  \caption{
    Conceptual diagram of curation and citation. We define curation as the transition from a publicly posted preprint to its subsequent publication in a peer-reviewed journal. Preprints that undergo this transition (curated preprints) provide the baseline for citation behavior, while preprints that remain unpublished (non-curated preprints) allow us to quantify the extent to which journals cite work outside the curated scholarly record.
 }
  \label{fig:curation_diagram}
\end{figure}
Curation is central to our analysis because it marks the point at which the work has passed formal peer review and thus moved from the early-dissemination stage to an established part of the scholarly record.
In this framework, preprints that later appear in peer-reviewed journals provide a natural baseline for citation behavior. When a curated journal version exists, it typically becomes the canonical object of citation. We therefore treat citations to curated preprints as a reference point and measure the extent to which journals cite preprints that remain non-curated.
Non-curated preprints are those that have not been published in any journal for at least two years after their initial posting.
This is a conventional time window given the fact that at least 90\% of preprints are curated within a year \parencite{AbdillBlekhman_TrackingPopularityOutcomes_2019}.

Curated preprints are initially identified using the metadata provided by the preprint servers themselves.
Specifically, we use the “doi” field in arXiv, the “published” field in bioRxiv and medRxiv, and the “attributes.doi” field in OSF Preprints, SocArXiv, and PsyArXiv metadata.
However, it is known that such server-level metadata can be incomplete.
Although most DOI-registered scholarly articles are indexed through Crossref, the linkage between preprints and their corresponding journal publications is not systematically guaranteed. These links depend on metadata deposited by publishers or repository maintainers. Because DOI registration for preprints only began to expand after 2016\parencite{RachaelLammey_PreprintsAreGo_2016}, and previous audits have reported missing links between preprints and their published versions\parencite{LinEtAl_HowManyPreprints_2020,CabanacEtAl_DaytodayDiscoveryPreprint_2021}, server-level metadata alone may fail to capture all curated preprints.
Consequently, relying solely on repository metadata may underestimate the true number of curated preprints.

To address this limitation, we performed an additional matching procedure to identify journal publications corresponding to preprints, following the general approach proposed by \cite{CabanacEtAl_DaytodayDiscoveryPreprint_2021}.
Two records were considered a match when the first author was identical and the Dice similarity between bag-of-words representations of the titles exceeded 0.3.
The similarity threshold was intentionally set higher than that used in the previous study. A lower threshold increases the risk of falsely identifying curated preprints, which would mechanically reduce the number of non-curated preprints in our dataset. 
Because our analysis compares citation rates to curated and non-curated preprints relative to a null model based on the growth of the preprint corpus, such false positives would bias the null model downward and artificially inflate the apparent citation rate of non-curated preprints. Adopting a stricter threshold therefore provides a conservative estimate that avoids overstating the prevalence of citations to non-curated preprints.
Using this procedure, we identified an additional 556,639 curated preprints across all six preprint servers. 
Importantly, only a fraction of these newly matched preprints were cited by journal articles in our dataset. 
Consequently, including matches with Dice similarity between 0.1 and 0.3 would have a negligible effect on the estimated citation rates.
Our conclusions therefore do not rely on borderline matches.

The resulting curation ratios become substantially higher than those obtained from server metadata alone.
For example, raw metadata indicates that 46.5\% of arXiv preprints, 57.1\% of bioRxiv preprints, and 52.9\% of medRxiv preprints are curated. After enrichment, these ratios increase markedly (Table 1), bringing them closer to previous estimates reported in the literature, such as \cite{LariviereEtAl_ArXivEprintsJournal_2014} and \cite{AbdillBlekhman_TrackingPopularityOutcomes_2019}.

\begin{table}[h]
  \centering
  \caption{Preprint statistics by preprint servers}
  \label{tab:preprint_matching}
  \begin{tabular}{lrrrrr}
    \toprule
    Preprint Server & Inception & Total & Curated (Ratio*1 (\%)) & Matched & Matching Rate\\
    \midrule
    arXiv         & 1991 & 2,633,433 & 1,635,472 (69.6\%) & 2,565,368 & 97.4\%  \\
    BioRxiv       & 2013 & 262,741 & 190,450 (78.5\%)& 262,598 & 99.9\%  \\
    OSF Preprints & 2016 & 70,518 & 11,872 (28.7\%)& 69,294 & 98.2\%  \\
    PsyArXiv      & 2016 & 39,782 & 18,298 (50.9\%)& 39,311 & 98.8\%  \\
    SocArXiv      & 2016 & 16,193 & 7,660 (48.7\%)& 16,114 & 99.5\%  \\
    medRxiv       & 2019 & 62,019 & 41,526 (78.7\%)& 61,973 & 99.9\%  \\
    \bottomrule
  \end{tabular}
  \begin{flushleft}
  \footnotesize
  *1 curated preprint ratio is calculated as the mean of yearly share of curated preprints until the end of 2022, considering a 2-year lag for curation. Note that the ratio does not match the ratio of curated preprints in total.
  \end{flushleft}
\end{table}

\subsection{Journal selection}
In this study, we focus on the citations from long-established journals with consistently rigorous editorial processes.
Our aim is to compare the role of preprints across multiple years.
To establish a baseline, we use citations to curated preprints and contrast them with citations to non-curated preprints. 
This allows us to separate the effect of accelerated research cycles and to assess the extent to which researchers rely on knowledge available only in preprints.
To achieve this, we require a stable population of journal articles each year and editorial policies that remain consistent over time. 
In addition, accurately identifying citations to curated preprints requires distinguishing between two cases: citations made before the journal article is formally published, and citations that still refer to the preprint even though the journal article is already available, the latter of which must be excluded from our analysis.
Because the recorded publication date of a journal article can shift during the editorial and production process, we use the editorial workflow itself as the basis for determining whether a citation genuinely predates the published article, instead of relying on nominal timestamps.
This distinction is dependable only when the editorial process is consistently enforced. 
Restricting our analysis to journals with dedicated, in-house editors and rigorous editorial policies ensures that these conditions are met and enhances the reliability of longitudinal comparisons.

\subsection{Null model for preprint citation}
Previous literature shows that the average number of citations to non-curated preprints is not fewer than that to curated preprints \parencite{FraserEtAl_RelationshipBioRxivPreprints_2020}. 
However, the mean number of citations per preprint is not an appropriate measure of citation preference, even on a logarithmic scale.
This is because the null expectation for preprint citation depends not on the absolute number of preprints but on their proportion among all citable items in a given year: if preprints occupy a growing share of the literature, higher citation counts may arise purely by chance.

To assess whether preprints are cited more or less than this null expectation, we constructed a reference set of citable items. We compiled the list of distinct venues cited by our journal set during the analysis timeframe: for journal articles, each journal title constitutes a venue; for preprints, we used subject-area categories as the unit of classification. We retained only venues cited at least five times between 2010 and 2024. Within this scope, we calculated, for each year, the proportion of preprints among all citable items (preprints and journal articles) and treated this proportion as the expected citation probability under the null model. Non-curated and curated preprints were evaluated separately. The observed citation rate was then computed as the proportion of all references from our citing journal set that pointed to preprints.

The ratio of observed to expected citation rates yields a likelihood ratio.
A ratio of 1 indicates that preprints are cited at exactly the rate predicted by their share of the literature.
A ratio above 1 suggests that researchers actively prefer to cite preprints, potentially owing to the quality of the work, the speed of access, or disciplinary citing norms, while a ratio below 1 implies the opposite.
Note that articles published in a given venue tend to preferentially cite other articles from the same venue, which could bias a likelihood ratio computed between journals. In our analysis, however, we measure citations from journal articles to preprints; because citations from preprints to preprints are not counted, this self-citation bias does not arise.

\section{Results}
\begin{figure}[H]
  \centering
  \includegraphics[width=0.75\linewidth]{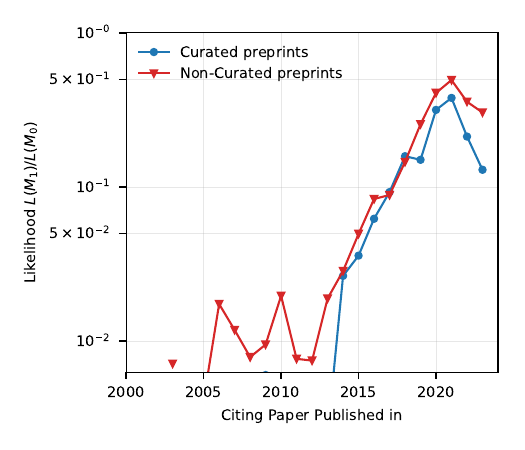}
  \caption{
    we compare the observed citation likelihood to a null model that reflects the expected citation likelihood if authors were indifferent between preprints and other forms of literature.
  }
  \label{fig:preprint_citation_null_model}
\end{figure}

Rather than examining the average impact of preprints, we quantified whether peer-review status produces a substantial difference in researchers' citation preferences, and if so, how this difference has evolved over time, using the likelihood ratio of citation probability relative to a null model.
Methodological details are provided in Section~\ref{sec:methods_preprint_curation}.
The citation likelihood of preprints is 10 to 20 times smaller than their share of the total literature under the null model (Figure~\ref{fig:preprint_citation_null_model}).
Nevertheless, journal articles have become increasingly likely to cite preprints, and this growth has been robust since 2014.
The rising trend is observed for both curated and non-curated preprints.
During the decade following the establishment of the q-bio category on arXiv in 2003, the citation likelihood of preprints remained largely unchanged.
Although citations to curated preprints existed before 2014, they numbered only a handful, yielding likelihood ratios below $10^{-3}$---essentially indistinguishable from noise.
Because the number of preprints has grown faster than the number of journal articles, the preprint share has increased consistently since 2009, with the sole exceptions being a decline for non-curated preprints between 2018 and 2019 and for curated preprints between 2021 and 2022.

During the COVID-19 pandemic period (2020--2022), the rise in the likelihood ratio of preprint citations stalled and subsequently declined. Further examination reveals that this reversal is attributed more to a deceleration in the growth of citation counts than to an increase in the preprint share of the literature.
Notably, citation counts for curated preprints had already been decelerating between 2018 and 2019, but surged sharply in 2020.
No comparable pattern is observed in the citation likelihood of non-curated preprints.

\begin{figure}[H]
  \centering
  \includegraphics[width=0.75\linewidth]{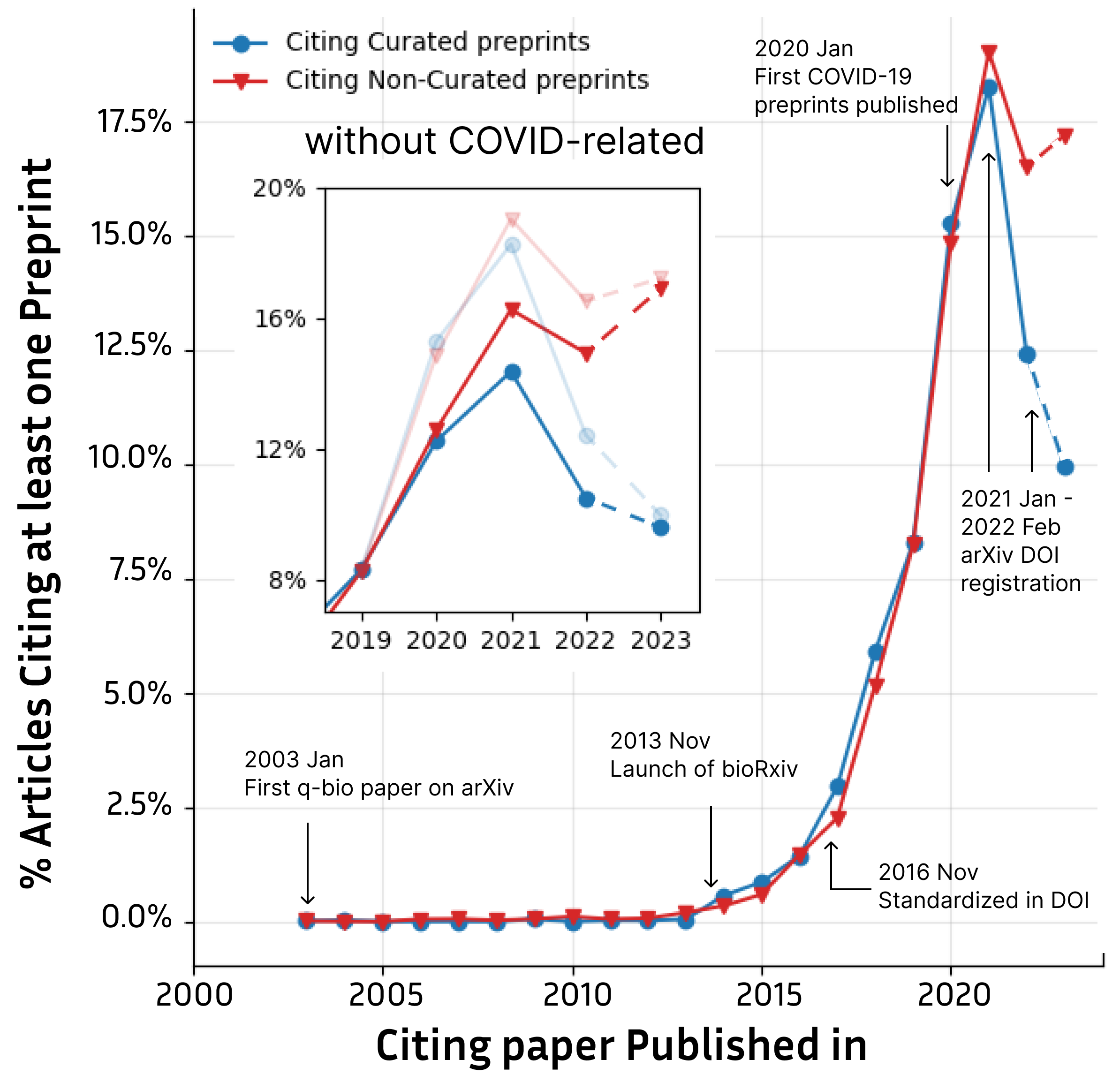}
  \caption{
  The fraction of articles citing one of the curated or non-curated preprints, grouped by the publication year of the citing journal article. Definition of curated and non-curated preprints are explained in the section \ref{sec:methods_preprint_curation}.
 }
  \label{fig:citing_rate_increase}
\end{figure}

It is possible that a limited fraction of articles in a subfield that frequently cite preprints are driving the overall increase in the likelihood ratio. To address this concern, we also examined the breadth of citation practices. Figure~\ref{fig:citing_rate_increase} shows that the fraction of journal articles citing preprints has increased exponentially since 2013.
The rising trend is observed for both curated and non-curated preprints, indicating that researchers draw on preprints regardless of whether they eventually appear in peer-reviewed journals. At its peak, more than 18\% of articles in our journal set cited at least one curated preprint, and nearly 19\% cited at least one non-curated preprint.
It is worth noting that the curated preprint citation ratio shows a modest decline in the most recent years (2022--2023). This pattern should be interpreted with caution, as curation typically involves a lag of approximately two years: preprints posted in recent years may not yet have appeared as journal articles, and are therefore still classified as non-curated at the time of observation.

Because the COVID-19 pandemic triggered an exceptional surge of preprint activity and accelerated journal publication in biomedical fields, we additionally examine citation trends after excluding journal articles explicitly related to COVID-19 (details in the section \ref{data_acquisition}).
During the early phase of the pandemic, the number of COVID-19 preprints and their subsequent journal publications increased sharply, with the surge beginning in early 2020 and stabilizing by late 2020 \parencite{AlgaEtAl_DevelopmentPreprintsCOVID19_2021}.
To ensure that the observed trends are not driven by this extraordinary period, we remove COVID-related citing journal articles from the analysis, while retaining all preprints in the dataset. 
As shown in the inset of Figure~\ref{fig:citing_rate_increase}, the overall increase in preprint citation remains clearly visible after this exclusion. 
Moreover, the gap between citations to non-curated and curated preprints becomes more pronounced once COVID-related articles are removed, indicating that the growing reliance on non-curated preprints is not simply a byproduct of pandemic-related research.

\begin{figure}[H]
  \centering
  \includegraphics[width=\linewidth]{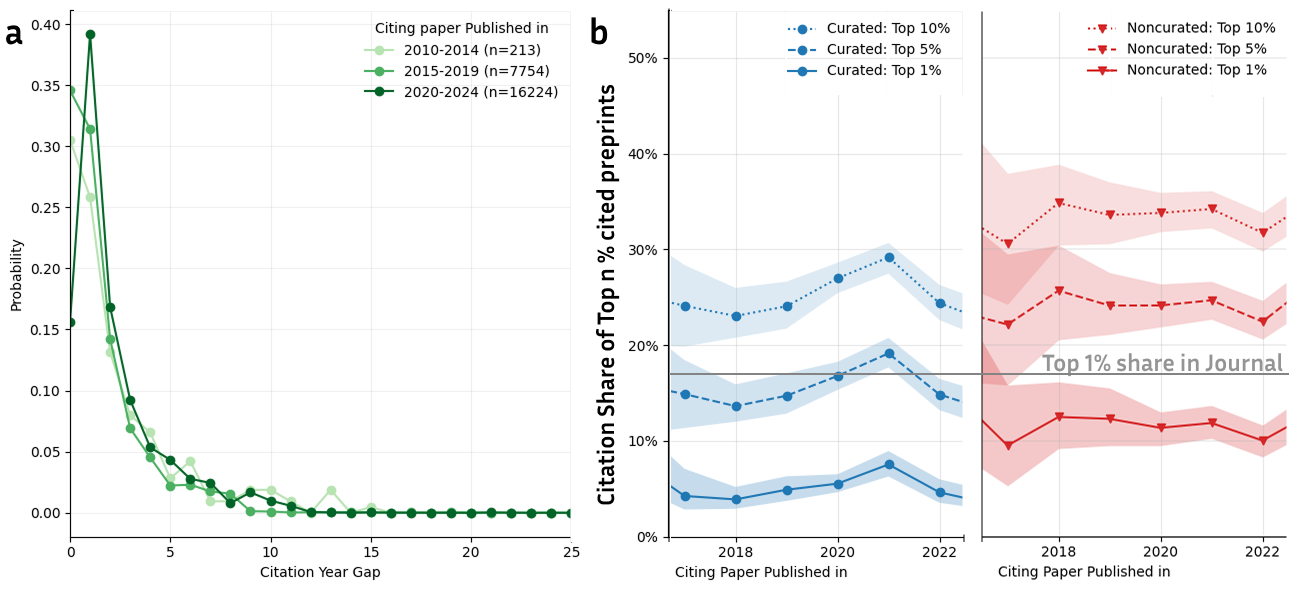}
  \caption{
    a) The normalized citation counts to preprints, deviding the total count of citations by the number of preprints available to cite, displayed for both curated and non-curated.
    b) Median citation top n\% share (n=1,5,10) for both curated (blue, circle markers) and non-curated(red, triangle markers) preprints. Bootstrapped 95\% confidence intervals are shown as shaded areas.
 }
  \label{fig:preprint_other}
\end{figure}

Another possibility is that accelerated research cycles have shortened the time between a preprint’s publication and its citation, leading to an artificial inflation of citation counts. 
However, the citation-year gap, namely the time between the preprint’s online publication and the citing journal article, remains remarkably stable across decades (Figure \ref{fig:preprint_other}a). The median citation lag is consistently around one and half to two years, and the overall distribution shows no structural shift between 2010–2015, 2016–2019, and 2020–2024. 
The mean citation-year gap from journal articles ($\pm 95\% $ CI) was approximately 2.33 (1.93 - 2.73), 1.55 (1.51 - 1.60), and 2.18 (2.15 - 2.22) years for 2010-15, 2016-19, and 2020-24, respectively.
When the aggregation is performed using left-closed intervals (2010--2014, 2015--2019, 2020--2024) rather than the right-closed intervals above, the 2010--2014 period emerges as an outlier with a wider mean gap, likely reflecting the small and heterogeneous early sample.

A further alternative is that a few highly influential preprints generate a disproportionate share of citations, producing the apparent generalized trend. 
If the rise in citations were driven by a small number of canonical preprints, top cited preprints would account for an increasing share of total citations over time.
We calculated top n\% (n=1, 5, 10) share to assess the temporal trend.

For both curated and non-curated preprints, we found a stable trend of citation share(Figure \ref{fig:preprint_other}b).
The top 1\% of most cited preprints consistently accounted for around 10\% (5\% for curated, same below) of all citations each year since 2017, while the top 5\% and 10\% accounted for approximately 25\% (15\%) and 35\% (25\%), respectively.
In early years before 2016, these share of preprints accounted for a higher share (up to 50\% in curated and up to 40\% in non-curated) of citations, but confidence intervals for the top-share estimates overlap substantially owing to the small number of cited preprints, limiting the reliability of early-period comparisons and thus we omitted from the visualization. For curated preprints, the top-share estimates steadily declined until 2017, and then stabilized at much lower levels in subsequent years, while for non-curated preprints, the top-share estimates has already stabilized since 2014, and median top 1\% share has been around 20\% in 2014.

Curated and non-curated preprints exhibited similar patterns, while non-curated preprints showed higher concentration across every comparison of the top $n$\% share we examined.
Note that even the concentration in non-curated preprints is comparably smaller than that of journal articles, where the top 1\% of articles account for around 17\% of citations \parencite{BarabasiEtAl_HandfulPapersDominates_2012, DongEtAl_CenturyScienceGlobalization_2017} at 2010.

During the COVID-19 period, the top share for curated preprints increased transiently, likely reflecting the outsized influence of a small number of heavily cited pandemic-related preprints that were rapidly curated. 
This pattern is also evident in the inset of Figure~\ref{fig:citing_rate_increase}. Between 2020 and 2022, when COVID-19-related journal articles are excluded, the decline is more pronounced for curated preprints.
In 2022, both curated and non-curated top shares declined simultaneously, coinciding with the normalization of pandemic-related preprint activity.

\section{Discussion}\label{sec:discussion}
Journals are not entirely indifferent to citing preprints, yet preprint citations remain a distinct minority compared with citations to journal articles, ten to twenty times smaller than expected by chance.
Nevertheless, the degree of preference for preprints---or at least the diminishing resistance to citing them---has increased at an exponential rate since the mid-2010s.
The share of journal articles that cite preprints has also grown, with preprint citations becoming broadly distributed across the biological and medical sciences.
At its peak in 2021, approximately 18\% of journal articles---nearly one in five---cited at least one preprint.
The growing preference for preprints is evident along two dimensions: not only in citation intensity but also in the breadth of adoption.

A factor that appears to have influenced both the intensity and adoption of citing non-curated preprints is the introduction of a dedicated preprint content type in Crossref's DOI system at the end of 2016.
Between 2017 and 2018, the preference for preprints strengthened discontinuously.
Because bioRxiv had already been assigning DOIs to preprints before 2016, the late-2016 institutional change did not merely improve the visibility of individual preprints but rather established preprints as a recognized bibliographic category within the scholarly infrastructure.
In contrast, a factor more likely to have driven both the intensity and adoption of citing curated preprints is the emergence of discipline-specific preprint servers.
Following the launch of bioRxiv in late 2013, the preference for preprints increased more than tenfold between 2013 and 2014.
Importantly, these patterns cannot be attributed simply to improved discoverability.
If discoverability were the primary driver, the expanding assignment of DOIs to arXiv preprints between 2021 and 2022, which enhanced their indexing in databases such as OpenAlex, should have led to an increase in preprint citations during that period.

Parallel to this trend, journals have gradually begun to cite older preprints: over the past decade, the mean gap between the publication year of the citing article and that of the cited preprint has widened by approximately one year.
By contrast, peer review duration has remained largely unchanged over the same period~\parencite{Publons_PublonsGlobalState_2018}.
This trend rules out explanations based on heightened urgency, shortened publication timelines, or delayed curation. The increased citation of preprints therefore reflects a deeper behavioral change, not a mechanical consequence of research speed.
Importantly, this widening citation-year gap does not simply reflect the continued citation of a few well-known legacy preprints.
This is evidenced by the fact that the share of total preprint citations accounted for by the most highly cited preprints has remained stable over the past decade.
Furthermore, the gap has widened by only about one year over a decade. Cited preprints are continuously turning over, with newer preprints successively entering the citation record.

Contrary to expectations, the heavy reliance on preprint during the COVID-19 pandemic did not further amplify the preference for citing preprints.
Instead, the preference for preprint citations decelerated during the pandemic period, and by 2022 the likelihood ratio had fallen to roughly half of its 2021 level.
A closer decomposition of the likelihood ratio reveals that the primary driver was a slowdown in the growth of preprint citation counts rather than a sharp increase in the article share. During the pandemic, the number of journal articles in the biological and medical sciences also grew, reflecting expedited peer review in response to the public health emergency~\parencite{Horbach_PandemicPublishingMedical_2020}.

Taken together, these patterns indicate that the rise in preprint citations is not an artifact of data growth, shorter publication cycles, or the increasing influence of a small number of highly visible manuscripts. Instead, they reveal a structural shift in citing behavior: preprints have become intentional, legitimate objects of citation, independent of their later publication status, with non-curated preprints cited at rates comparable to curated ones. Crucially, however, our findings suggest that the decision to cite a preprint depends less on the relevance of its content and more on the degree to which the preprint medium itself is trusted. The formal recognition of preprints as a dedicated content type in the Crossref DOI system, an institutional signal of legitimacy, was followed by a discontinuous increase in citation preference. Conversely, the COVID-19 pandemic, despite producing a surge in preprint volume and absolute citation counts, coincided with a decline in preference. The pandemic experience actively eroded trust: comparisons of COVID-19 preprints with their published journal versions revealed substantial discrepancies in results and interpretation~\parencite{BeroEtAl_ComparisonPreprintsFinal_2021}, and nearly half of medRxiv preprints cited in leading medical journals differed in title, data, or conclusions from their final form~\parencite{GehannoEtAl_ReliabilityCitationsMedRxiv_2022}, prompting calls for automated screening to improve transparency and reproducibility~\parencite{WeissgerberEtAl_AutomatedScreeningCOVID19_2021}. Researcher surveys confirm that low reliability and credibility remain the primary concerns discouraging preprint use, particularly in the life and health sciences~\parencite{NiWaltman_PreprintNotPreprint_2024,FraserEtAl_MotivationsConcernsSelection_2022}.

These results imply that the future of preprints as an integral component of the scientific communication system hinges on sustained community trust\parencite{KlingMcKim_NotJustMatter_2000}. The challenges are already visible: in machine learning, major venues have introduced restrictions on position papers to curb declining submission quality, and calls for robust governance of individual preprint servers are intensifying~\parencite{_AttentionAuthorsUpdated_b}. Whether preprints consolidate their role as autonomous, citable sources of knowledge will depend on the capacity of servers and communities to uphold standards of rigor that justify the trust researchers place in them.

\section*{Data Availability}
The data and code will be publicly available at Zenodo.
DOI: 10.5281/zenodo.19324412

\printbibliography

@online{_DirectoryOpenAccess_,
  title    = {Directory of {{Open Access Preprint Repositories}}: {{Repositories}}},
  url      = {https://doapr.coar-repositories.org/repositories/},
  urldate  = {2025-12-23}
}

@article{AbdillBlekhman_TrackingPopularityOutcomes_2019,
  title        = {Tracking the Popularity and Outcomes of All {{bioRxiv}} Preprints},
  author       = {Abdill, Richard J and Blekhman, Ran},
  editor       = {Pewsey, Emma and Rodgers, Peter and Greene, Casey S},
  date         = {2019-04-24},
  journaltitle = {eLife},
  volume       = {8},
  pages        = {e45133},
  publisher    = {eLife Sciences Publications, Ltd},
  issn         = {2050-084X},
  doi          = {10.7554/eLife.45133},
  url          = {https://doi.org/10.7554/eLife.45133},
  urldate      = {2025-01-15}
}

@article{AkbaritabarEtAl_StudyReferencingChanges_2022,
  title        = {A Study of Referencing Changes in Preprint-Publication Pairs across Multiple Fields},
  author       = {Akbaritabar, Aliakbar and Stephen, Dimity and Squazzoni, Flaminio},
  date         = {2022-05-01},
  journaltitle = {Journal of Informetrics},
  shortjournal = {Journal of Informetrics},
  volume       = {16},
  number       = {2},
  pages        = {101258},
  issn         = {1751-1577},
  doi          = {10.1016/j.joi.2022.101258},
  url          = {https://www.sciencedirect.com/science/article/pii/S1751157722000104},
  urldate      = {2025-01-23}
}

@article{Amelino-Camelia_QuantumTheorysLast_2000,
  title        = {Quantum Theory's Last Challenge},
  author       = {Amelino-Camelia, Giovanni},
  date         = {2000-12},
  journaltitle = {Nature},
  volume       = {408},
  number       = {6813},
  pages        = {661--664},
  publisher    = {Nature Publishing Group},
  issn         = {1476-4687},
  doi          = {10.1038/35047210},
  url          = {https://www.nature.com/articles/35047210},
  urldate      = {2025-11-06},
  langid       = {english}
}

@article{BergEtAl_PreprintsLifeSciences_2016,
  title        = {Preprints for the Life Sciences},
  author       = {Berg, Jeremy M. and Bhalla, Needhi and Bourne, Philip E. and Chalfie, Martin and Drubin, David G. and Fraser, James S. and Greider, Carol W. and Hendricks, Michael and Jones, Chonnettia and Kiley, Robert and King, Susan and Kirschner, Marc W. and Krumholz, Harlan M. and Lehmann, Ruth and Leptin, Maria and Pulverer, Bernd and Rosenzweig, Brooke and Spiro, John E. and Stebbins, Michael and Strasser, Carly and Swaminathan, Sowmya and Turner, Paul and Vale, Ronald D. and VijayRaghavan, K. and Wolberger, Cynthia},
  date         = {2016-05-20},
  journaltitle = {Science},
  volume       = {352},
  number       = {6288},
  pages        = {899--901},
  publisher    = {American Association for the Advancement of Science},
  doi          = {10.1126/science.aaf9133},
  url          = {https://www.science.org/doi/full/10.1126/science.aaf9133},
  urldate      = {2025-01-13}
}

@online{BeroEtAl_ComparisonPreprintsFinal_2021,
  title      = {Comparison of Preprints and Final Journal Publications from {{COVID-19 Studies}}: {{Discrepancies}} in Results Reporting and Spin in Interpretation},
  shorttitle = {Comparison of Preprints and Final Journal Publications from {{COVID-19 Studies}}},
  author     = {Bero, Lisa and Lawrence, Rosa and Leslie, Louis and Chiu, Kellia and McDonald, Sally and Page, Matthew J. and Grundy, Quinn and Parker, Lisa and Boughton, Stephanie L. and Kirkham, Jamie J. and Featherstone, Robin},
  date       = {2021-04-19},
  eprinttype = {medRxiv},
  pages      = {2021.04.12.21255329},
  doi        = {10.1101/2021.04.12.21255329},
  url        = {https://www.medrxiv.org/content/10.1101/2021.04.12.21255329v1},
  urldate    = {2025-12-23},
  langid     = {english},
  pubstate   = {prepublished}
}

@article{Bjork_OpenAccessSubject_2014,
  title        = {Open Access Subject Repositories: {{An}} Overview},
  shorttitle   = {Open Access Subject Repositories},
  author       = {Björk, Bo-Christer},
  date         = {2014},
  journaltitle = {Journal of the Association for Information Science and Technology},
  volume       = {65},
  number       = {4},
  pages        = {698--706},
  issn         = {2330-1643},
  doi          = {10.1002/asi.23021},
  url          = {https://onlinelibrary.wiley.com/doi/abs/10.1002/asi.23021},
  urldate      = {2025-12-31},
  langid       = {english}
}

@article{CabanacEtAl_DaytodayDiscoveryPreprint_2021,
  title        = {Day-to-Day Discovery of Preprint–Publication Links},
  author       = {Cabanac, Guillaume and Oikonomidi, Theodora and Boutron, Isabelle},
  date         = {2021-06-01},
  journaltitle = {Scientometrics},
  shortjournal = {Scientometrics},
  volume       = {126},
  number       = {6},
  pages        = {5285--5304},
  issn         = {1588-2861},
  doi          = {10.1007/s11192-021-03900-7},
  url          = {https://doi.org/10.1007/s11192-021-03900-7},
  urldate      = {2025-03-08},
  langid       = {english}
}

@article{EisenEtAl_ImplementingPublishThen_2020,
  title        = {Implementing a "Publish, Then Review" Model of Publishing},
  author       = {Eisen, Michael B and Akhmanova, Anna and Behrens, Timothy E and Harper, Diane M and Weigel, Detlef and Zaidi, Mone},
  date         = {2020-12-01},
  journaltitle = {eLife},
  volume       = {9},
  pages        = {e64910},
  issn         = {2050-084X},
  doi          = {10.7554/eLife.64910},
  url          = {https://elifesciences.org/articles/64910},
  urldate      = {2025-01-14},
  langid       = {english}
}

@article{FraserEtAl_MotivationsConcernsSelection_2022,
  title        = {Motivations, Concerns and Selection Biases When Posting Preprints: {{A}} Survey of {{bioRxiv}} Authors},
  shorttitle   = {Motivations, Concerns and Selection Biases When Posting Preprints},
  author       = {Fraser, Nicholas and Mayr, Philipp and Peters, Isabella},
  date         = {2022-11-03},
  journaltitle = {PLOS ONE},
  shortjournal = {PLOS ONE},
  volume       = {17},
  number       = {11},
  pages        = {e0274441},
  publisher    = {Public Library of Science},
  issn         = {1932-6203},
  doi          = {10.1371/journal.pone.0274441},
  url          = {https://journals.plos.org/plosone/article?id=10.1371/journal.pone.0274441},
  urldate      = {2025-01-28},
  langid       = {english}
}

@article{GehannoEtAl_ReliabilityCitationsMedRxiv_2022,
  title        = {Reliability of Citations of {{medRxiv}} Preprints in Articles Published on {{COVID-19}} in the World Leading Medical Journals},
  author       = {Gehanno, Jean-Francois and Grosjean, Julien and Darmoni, Stefan J. and Rollin, Laetitia},
  date         = {2022-08-10},
  journaltitle = {PLOS ONE},
  shortjournal = {PLOS ONE},
  volume       = {17},
  number       = {8},
  pages        = {e0264661},
  publisher    = {Public Library of Science},
  issn         = {1932-6203},
  doi          = {10.1371/journal.pone.0264661},
  url          = {https://journals.plos.org/plosone/article?id=10.1371/journal.pone.0264661},
  urldate      = {2025-09-22},
  langid       = {english}
}

@dataset{Geiger_ArXiVArchive_2019,
  title     = {{{ArXiV Archive}}},
  author    = {Geiger, R.Stuart},
  date      = {2019-01-01},
  publisher = {Zenodo},
  doi       = {10.6078/D1708G},
  url       = {https://zenodo.org/records/4990937},
  urldate   = {2025-01-13}
}

@article{Goldschmidt-Clermont_CommunicationPatternsHighEnergy_2002,
  entrysubtype = {magazine},
  title        = {Communication {{Patterns}} in {{High-Energy Physics}}},
  author       = {Goldschmidt-Clermont, Luisella},
  date         = {2002-03},
  journaltitle = {High Energy Physics Libraries Webzine},
  number       = {6},
  url          = {https://cds.cern.ch/record/546422/files/sis-2002-163.html},
  urldate      = {2025-01-30},
  langid       = {english}
}

@article{Horbach_PandemicPublishingMedical_2020,
  title        = {Pandemic Publishing: {{Medical}} Journals Strongly Speed up Their Publication Process for {{COVID-19}}},
  shorttitle   = {Pandemic Publishing},
  author       = {Horbach, Serge P. J. M.},
  date         = {2020-08-01},
  journaltitle = {Quantitative Science Studies},
  shortjournal = {Quantitative Science Studies},
  volume       = {1},
  number       = {3},
  pages        = {1056--1067},
  issn         = {2641-3337},
  doi          = {10.1162/qss_a_00076},
  url          = {https://doi.org/10.1162/qss_a_00076},
  urldate      = {2025-01-15}
}

@article{Kelly_SIGIRCommunitySurvey_2018,
  title        = {{{SIGIR Community Survey}} on {{Preprint Services}}},
  author       = {Kelly, Diane},
  date         = {2018-08-31},
  journaltitle = {SIGIR Forum},
  volume       = {52},
  number       = {1},
  pages        = {11--33},
  issn         = {0163-5840},
  doi          = {10.1145/3274784.3274787},
  url          = {https://dl.acm.org/doi/10.1145/3274784.3274787},
  urldate      = {2025-11-21}
}

@article{KurtzEtAl_EffectUseAccess_2005,
  title        = {The Effect of Use and Access on Citations},
  author       = {Kurtz, Michael J. and Eichhorn, Guenther and Accomazzi, Alberto and Grant, Carolyn and Demleitner, Markus and Henneken, Edwin and Murray, Stephen S.},
  date         = {2005-12},
  journaltitle = {Information Processing \& Management},
  shortjournal = {Information Processing \& Management},
  volume       = {41},
  number       = {6},
  pages        = {1395--1402},
  issn         = {03064573},
  doi          = {10.1016/j.ipm.2005.03.010},
  url          = {https://linkinghub.elsevier.com/retrieve/pii/S0306457305000361},
  urldate      = {2025-11-11},
  langid       = {english}
}

@article{LariviereEtAl_ArXivEprintsJournal_2014,
  title        = {{{arXiv E-prints}} and the Journal of Record: {{An}} Analysis of Roles and Relationships},
  shorttitle   = {{{arXiv E-prints}} and the Journal of Record},
  author       = {Larivière, Vincent and Sugimoto, Cassidy R. and Macaluso, Benoit and Milojević, Staša and Cronin, Blaise and Thelwall, Mike},
  date         = {2014},
  journaltitle = {Journal of the Association for Information Science and Technology},
  volume       = {65},
  number       = {6},
  pages        = {1157--1169},
  issn         = {2330-1643},
  doi          = {10.1002/asi.23044},
  url          = {https://onlinelibrary.wiley.com/doi/abs/10.1002/asi.23044},
  urldate      = {2025-06-06},
  langid       = {english}
}

@article{LinEtAl_HowManyPreprints_2020,
  title        = {How Many Preprints Have Actually Been Printed and Why: A Case Study of Computer Science Preprints on {{arXiv}}},
  shorttitle   = {How Many Preprints Have Actually Been Printed and Why},
  author       = {Lin, Jialiang and Yu, Yao and Zhou, Yu and Zhou, Zhiyang and Shi, Xiaodong},
  date         = {2020-07-01},
  journaltitle = {Scientometrics},
  shortjournal = {Scientometrics},
  volume       = {124},
  number       = {1},
  pages        = {555--574},
  issn         = {1588-2861},
  doi          = {10.1007/s11192-020-03430-8},
  url          = {https://doi.org/10.1007/s11192-020-03430-8},
  urldate      = {2025-09-20},
  langid       = {english}
}

@article{NiWaltman_PreprintNotPreprint_2024,
  title        = {To Preprint or Not to Preprint: {{A}} Global Researcher Survey},
  shorttitle   = {To Preprint or Not to Preprint},
  author       = {Ni, Rong and Waltman, Ludo},
  date         = {2024},
  journaltitle = {Journal of the Association for Information Science and Technology},
  volume       = {75},
  number       = {6},
  pages        = {749--766},
  issn         = {2330-1643},
  doi          = {10.1002/asi.24880},
  url          = {https://onlinelibrary.wiley.com/doi/abs/10.1002/asi.24880},
  urldate      = {2025-02-26},
  langid       = {english}
}

@article{AlgaEtAl_DevelopmentPreprintsCOVID19_2021,
  title        = {The Development of Preprints during the {{COVID-19}} Pandemic},
  author       = {Älgå, Andreas and Eriksson, Oskar and Nordberg, Martin},
  date         = {2021},
  journaltitle = {Journal of Internal Medicine},
  volume       = {290},
  number       = {2},
  pages        = {480--483},
  issn         = {1365-2796},
  doi          = {10.1111/joim.13240},
  url          = {https://onlinelibrary.wiley.com/doi/abs/10.1111/joim.13240},
  urldate      = {2025-11-03},
  langid       = {english}
}

@article{DongEtAl_CenturyScienceGlobalization_2017,
  title         = {A {{Century}} of {{Science}}: {{Globalization}} of {{Scientific Collaborations}}, {{Citations}}, and {{Innovations}}},
  shorttitle    = {A {{Century}} of {{Science}}},
  author        = {Dong, Yuxiao and Ma, Hao and Shen, Zhihong and Wang, Kuansan},
  year          = 2017,
  month         = oct,
  number        = {arXiv:1704.05150},
  eprint        = {1704.05150},
  primaryclass  = {cs},
  publisher     = {arXiv},
  doi           = {10.48550/arXiv.1704.05150},
  urldate       = {2026-01-18},
  archiveprefix = {arXiv}
}

@online{RachaelLammey_PreprintsAreGo_2016,
  type         = {website},
  title        = {Preprints Are Go at {{Crossref}}!},
  author       = {{Rachael Lammey}},
  date         = {2016-11-02},
  doi          = {10.64000/5tcfp-vf140},
  url          = {https://www.crossref.org/blog/preprints-are-go-at-crossref/},
  urldate      = {2025-11-21},
  langid       = {english},
  organization = {Crossref}
}

@article{WeissgerberEtAl_AutomatedScreeningCOVID19_2021,
  title        = {Automated Screening of {{COVID-19}} Preprints: Can We Help Authors to Improve Transparency and Reproducibility?},
  shorttitle   = {Automated Screening of {{COVID-19}} Preprints},
  author       = {Weissgerber, Tracey and Riedel, Nico and Kilicoglu, Halil and Labbé, Cyril and Eckmann, Peter and family=Riet, given=Gerben, prefix=ter, useprefix=true and Byrne, Jennifer and Cabanac, Guillaume and Capes-Davis, Amanda and Favier, Bertrand and Saladi, Shyam and Grabitz, Peter and Bannach-Brown, Alexandra and Schulz, Robert and McCann, Sarah and Bernard, Rene and Bandrowski, Anita},
  date         = {2021-01},
  journaltitle = {Nature Medicine},
  shortjournal = {Nat Med},
  volume       = {27},
  number       = {1},
  pages        = {6--7},
  publisher    = {Nature Publishing Group},
  issn         = {1546-170X},
  doi          = {10.1038/s41591-020-01203-7},
  url          = {https://www.nature.com/articles/s41591-020-01203-7},
  urldate      = {2024-06-03},
  langid       = {english}
}

@article{BarabasiEtAl_HandfulPapersDominates_2012,
  title     = {Handful of Papers Dominates Citation},
  author    = {Barab{\'a}si, Albert-L{\'a}szl{\'o} and Song, Chaoming and Wang, Dashun},
  year      = 2012,
  month     = nov,
  journal   = {Nature},
  volume    = {491},
  number    = {7422},
  pages     = {40--40},
  publisher = {Nature Publishing Group},
  issn      = {1476-4687},
  doi       = {10.1038/491040a},
  urldate   = {2026-01-18},
  copyright = {2012 Springer Nature Limited},
  langid    = {english}
}

@article{FraserEtAl_RelationshipBioRxivPreprints_2020,
  title    = {The Relationship between {{bioRxiv}} Preprints, Citations and Altmetrics},
  author   = {Fraser, Nicholas and Momeni, Fakhri and Mayr, Philipp and Peters, Isabella},
  year     = 2020,
  month    = jun,
  journal  = {Quantitative Science Studies},
  volume   = {1},
  number   = {2},
  pages    = {618--638},
  issn     = {2641-3337},
  doi      = {10.1162/qss_a_00043},
  urldate  = {2025-01-15}
}

@article{KlingMcKim_NotJustMatter_2000,
  title        = {Not Just a Matter of Time: {{Field}} Differences and the Shaping of Electronic Media in Supporting Scientific Communication},
  shorttitle   = {Not Just a Matter of Time},
  author       = {Kling, Rob and McKim, Geoffrey},
  date         = {2000},
  journaltitle = {Journal of the American Society for Information Science},
  volume       = {51},
  number       = {14},
  pages        = {1306--1320},
  issn         = {1097-4571},
  doi          = {10.1002/1097-4571(2000)9999:9999<::AID-ASI1047>3.0.CO;2-T},
  url          = {https://onlinelibrary.wiley.com/doi/abs/10.1002/1097-4571%282000%299999%3A9999%3C%3A%3AAID-ASI1047%3E3.0.CO%3B2-T},
  urldate      = {2025-12-25},
  langid       = {english}
}

@online{_AttentionAuthorsUpdated_b,
  title    = {Attention {{Authors}}: {{Updated Practice}} for {{Review Articles}} and {{Position Papers}} in {{arXiv CS Category}} – {{arXiv}} Blog},
  url      = {https://blog.arxiv.org/2025/10/31/attention-authors-updated-practice-for-review-articles-and-position-papers-in-arxiv-cs-category/},
  urldate  = {2026-04-01}
}

@report{Publons_PublonsGlobalState_2018,
  title       = {Publons' {{Global State Of Peer Review}} 2018},
  author      = {{Publons}},
  date        = {2018-09-07},
  edition     = {0},
  institution = {Publons},
  location    = {London, UK},
  doi         = {10.14322/publons.GSPR2018},
  url         = {https://publons.com/static/Publons-Global-State-Of-Peer-Review-2018.pdf},
  urldate     = {2025-05-30},
  langid      = {english}
}

\end{document}